\documentclass[aps,prb,reprint,10pt,fleqn,showpacs,showkeys,floatfix,superscriptaddress,nofootinbib]{revtex4-1}
\usepackage{graphicx,color}
\usepackage{amssymb}
\usepackage{amsmath}
\usepackage{bm}
\usepackage{dcolumn}% Align table columns on decimal point
\usepackage[normalem]{ulem}
\begin{document}

\title{Magnetic Structure of the Mixed Antiferromagnet NdMn$_{0.8}$Fe$_{0.2}$O$_{3}$}

\author{Mat\'{u}\v{s} Mihalik}
\email{matmihalik@saske.sk}
\affiliation{Institute of Experimental Physics SAS, Watsonova 47, 040 01 Ko\v{s}ice, Slovak Republic}
\author{Mari\'{a}n Mihalik}
\affiliation{Institute of Experimental Physics SAS, Watsonova 47, 040 01 Ko\v{s}ice, Slovak Republic}
\author{Andreas Hoser}
\affiliation{Helmholtz-Zentrum Berlin, Hahn-Meitner-Platz 1, 14109 Berlin, Germany}
\author{Daniel M.~Pajerowski}
\email{pajerowskidm@ornl.gov}
\affiliation{Quantum Condensed Matter Division, Neutron Sciences Directorate, Oak Ridge National 
Laboratory, Oak Ridge, Tennessee 37831, USA}
\author{Dominik Kriegner}
\affiliation{Department of Condensed Matter Physics, Faculty of Mathematics and Physics, 
Charles University, Ke Karlovu 5, CZ-121 16 Praha 2, Czech Republic}
\author{Dominik Legut}
\affiliation{IT4Innovations Center, VSB-Technical University of Ostrava, 17. listopadu 15, CZ-708 33 Ostrava, Czech Republic}
\author{Kristof M.~Lebecki}
\affiliation{IT4Innovations Center, VSB-Technical University of Ostrava, 17. listopadu 15, CZ-708 33 Ostrava, Czech Republic}
\author{Martin Vavra}
\affiliation{Department of Inorganic Chemistry, Institute of Chemistry, Faculty of Science, 
P.~J.~\v{S}af\'{a}rik University, Moyzesova 11, 041 54 Ko\v{s}ice, Slovak Republic}
\author{Magdalena Fitta}
\affiliation{Institute of Nuclear Physics Polish Academy of Sciences, Radzikowskiego 152, 31-342 Krak\'{o}w, Poland}
\author{Mark~W.~Meisel}
\email{meisel@phys.ufl.edu}
\affiliation{Department of Physics and National High Magnetic Field Laboratory, University of Florida, 
Gainesville, FL 32611-8440, USA}
\affiliation{Joint Institute for Neutron Sciences, Oak Ridge National Laboratory, Oak Ridge, TN 37831-6453, USA}
\affiliation{Department of Condensed Matter Physics, Institute of Physics, Faculty of Science, 
P.~J.~\v{S}af\'{a}rik University, Park Angelinum 9, 041 54 Ko\v{s}ice, Slovak Republic}

\begin{abstract}
The magnetic structure of the mixed antiferromagnet NdMn$_{0.8}$Fe$_{0.2}$O$_3$ was resolved.    
Neutron powder diffraction data definitively resolve the Mn-sublattice with a magnetic propagation vector ${\bf k} = (000)$ and 
with the magnetic structure (A$_x$, F$_y$, G$_z$) for 1.6~K~$< T < T_N (\approx 59$~K). The Nd-sublattice has a (0, f$_y$, 0) contribution in the same temperature interval. The Mn sublattice undergoes spin-reorientation transition at $T_1 \approx 13$~K while the Nd magnetic moment keep ordered abruptly increases at this temperature. Powder X-ray diffraction shows a strong magnetoelastic effect at $T_N$ but no additional structural phase transitions from 2~K to 300~K.   Density functional theory calculations confirm the magnetic structure of the undoped NdMnO$_3$ as part of our analysis.  
Taken together, these results show the magnetic structure of Mn-sublattice in NdMn$_{0.8}$Fe$_{0.2}$O$_3$ is a combination of the Mn and Fe parent 
compounds, but the magnetic ordering of Nd sublattice spans over broader temperature interval than in case of NdMnO$_3$ and NdFeO$_3$. This result is a consequence of the fact that the Nd ions do not order independently, but via polarization from Mn/Fe sublattice.
\end{abstract}

\date{\today}% It is always \today, today, but any date may be explicitly specified


\maketitle

\section{Introduction}

Complex oxides, of which manganites are a subset, host multiferroicity and magnetoelectricity.\cite{Martin2012}  
One motivation for investigating $R_{1-x}A_x$MnO$_3$ hole-doped manganites, where $R$ is a rare earth and $A$ is an alkali element, 
is colossal magnetoresistance (CMR).\cite{Coey1998,Salamon2001}  The most studied CMR material is La$_{1-x}$Ca$_x$MnO$_3$ that shows a complex 
interplay between magnetic, charge, and structural order, all of which may affect CMR,\cite{Ramirez1997} and the Nd$_{1-x}$Ca$_x$MnO$_3$ material 
shows similar CMR features.\cite{Millange2000}  Recently, neutron scattering experiments and density functional theory analysis of 
SrMnO$_3$ and NdMnO$_3$ heterostructures displayed an interfacial ferromagnetism that is a step towards manganite-based 
multiferroic devices.\cite{Glavic2016} 

NdMn$_{1-x}$Fe$_x$O$_3$ is a magnetic insulator that contains three ions with well documented magnetochemistry.\cite{Carlin1986}   
The Nd$^{3+}$ has a 4$f^3$, 10-fold degenerate magnetic $^4$I ground state, which is split and mixed in the perovskite 
host lattice to have both orbital and spin components.  The Mn$^{3+}$ ion is $S = 2$, 3$d^4$ with a Jahn-Teller active $^5$E$_{\rm{g}}$ 
ground state.  The Fe$^{3+}$ ion has a half-full $d$-shell $S = 5/2$, $^6$A$_1$ ground state.  An understanding of the alloyed, 
solid-solution materials begins with a description of the well-studied end members NdMnO$_3$ and NdFeO$_3$.  
In the following, the discussion will be restricted to $T > T_{\rm{Nd}} \approx 1.5$~K), where the Nd-Nd interaction 
becomes important and an additional order parameter must be introduced.\cite{bartolome1997}

Neutron diffraction studies have shown that NdMnO$_3$ is an A-type antiferromagnet, where Mn sublattice orders to make it orders into a $\mu_{\rm{Mn}} = $~(A$_x$, F$_y$, 0)\cite{Munoz2000} or $\mu_{\rm{Mn}} = $~(A$_x$,0, 0) \cite{Chatterji2009} magnetic structure below $T_{\rm{N}} = 82$~K.\cite{Munoz2000,Chatterji2009} The moment axes are dictated by the strong anisotropy ($D \approx 5$~K) of the Jahn-Teller distorted manganese.\cite{Jahn1937}  In NdMnO$_3$, below $T_1 = 20$~K, there is a 
second transition that is associated with a ferromagnetic Mn-Nd interaction causing ordering of Nd sublattice to the  $\mu_{\rm{Nd}} = $~(0, f$_y$, 0)  magnetic structure,\cite{Munoz2000,Chatterji2009} while no effect on the Mn sublattice was observed\cite{Munoz2000} or additional canting of the Mn-moments to the $\mu_{\rm{Mn}} = $~(A$_x$, F$_y$, 0) is reported. \cite{Chatterji2009} It is noteworthy that, due to antisymmetric exchange, weak ferromagnetism in $R$MnO$_3$ compounds gives (A$_x$, F$_y$, 0) ordering with F$_y$ and/or A$_x \ll 1$ being rather small for the majority of light rare earth ions.\cite{Munoz2000,Chatterji2009,Skumryev1999,jirak1997,flynn2011} The magnetic excitation of the Nd ions below $T_1$ has been confirmed by a neutron backscattering experiment\cite{Chatterji2009bis} with no applied magnetic field, but the same experiment revealed non-zero polarization of the Nd ions below 40~K while X-ray magnetic circular dichroism data acquired in an applied magnetic field showed ordering of Nd sublattice  already below $T_{\rm N}$.\cite{Bartolome2005}

NdFeO$_3$ is a G-type antiferromagnet with weak ferromagnetism with $\mu_{\rm{Fe}}$~=~(0, F$_y$, G$_z$)\footnote{Unless otherwise stated, the data and analysis reported herein uses the orthorhombic space group $Pnma$.} and $T_N = 690$~K.\cite{Belov1972,Koshizuka1988}   
The Fe-octahedra in NdFeO$_3$ are nearly isotropic, in contrast to NdMnO$_3$, with a much weaker magnetic anisotropy 
($D < 0.1$~K).  Indeed, anisotropy in orthoferrites is subtle enough that spin reorientation due to octahedra rotation as a function of temperature 
is typical, and in NdFeO$_3$, the iron moments undergo a continuous transition to (0, G$_y$, F$_z$) from 167~K to 125~K concomitant with an 
octahedra rotation.\cite{slawinski2005}  For $T < 10$~K, the Nd-Fe interaction induces a noticeable Nd moment \cite{bartolome1997} that has a gradual onset into (c$_x$, 0, f$_z$) structure with the difference that the weak ferromagnetism of the two sublattices is antiparallel in NdFeO$_3$.\cite{slawinski2005,bartolome1997}  Neutron backscattering has also confirmed the Nd magnetic moment in NdFeO$_3$ for temperatures below 4.5~K.\cite{Przenioslo2006}

So in the context of the two pure end-point compounds, NdMn$_{1-x}$Fe$_x$O$_3$ is a mixed-anisotropy, mixed-type antiferromagnet with a phase diagram reported for $0 \leq x \leq 0.5$  that shows a similar suppression of $T_{\rm{N}}$ in the given $x$ interval as other members of the $R$Mn$_{1-x}$Fe$_x$O$_3$ family.\cite{Troyanchuk2007, chiang2011, mihalik2016, mihalik2017} Recently, additional investigations have been performed for  $0 \leq x \leq 0.3$.\cite{Mihalik2013,Mihalik2014,Lazurova2015} The substitution of Fe$^{3+}$ for Mn$^{3+}$ ions modifies the superexchange interactions, alters the polarization of the Nd$^{3+}$ ions through the Nd-Mn and Nd-Fe interactions, and changes the electron-phonon coupling due to reduction of the Jahn-Teller effect.\cite{Troyanchuk2001}  Although NdFeO$_3$ has a significantly higher ordering temperature 
than NdMnO$_3$, $T_{\rm{N}}$ is found to monotonically \emph{decrease} with iron doping in the range of $0 \leq x \leq 0.3$ that was studied.\cite{Mihalik2013}  
On the other hand, below $T_{\rm{N}}$, a low-temperature magnetic transition, $T_1$, defined by an anomaly in AC susceptibility decreases with increasing doping.\cite{Mihalik2013}  Extrapolating from the NdMnO$_3$ compound, this anomaly was tentatively assigned to the ordering of Nd ions,\cite{Mihalik2013} although no microscopic study of this transition has been published until now. The AC susceptibility peak width of $T_{\rm{N}}$ broadens with increasing $x$, for both in phase $(\chi^{\prime})$ and out-of-phase $(\chi^{\prime \prime})$ components.  The AC peak associated with $T_1$ varies non-monotonically in position between 11~K to 16~K and also in intensity, with the maximum intensity of $\chi^{\prime}$ and $\chi^{\prime \prime}$ when $x = 0.2$.  
Furthermore, hysteretic behavior between magnetization measurements with zero-field-cooled (ZFC) and field-cooled (FC) protocols was observed, 
while magnetic pole inversion, with a compensation temperature near 27~K was observed for samples with $x = 0.2$ and 0.25.\cite{Mihalik2013}  

The alloying of different type antiferromagnets has previously been studied in detail to understand the oblique antiferromagnetic (OAF) 
phase where spins point at an angle between those of the parent compounds, and three classic examples are the highly two-dimensional 
Fe$_{1-x}$Co$_x$Cl$_2$\cite{Tawaraya1980} and the three-dimensional, hexagonal K$_2$Mn$_{1-x}$Fe$_x$F$_4$\cite{Bevaart1978} and 
Fe$_x$Co$_{1-x}$TiO$_3$\cite{Torikai1986} compounds.  Such compounds show the characteristic dip in ordering temperatures between the 
two parent compounds with a minimum at a tricritical point in the $x - T$ plane, similar to $R$Mn$_{1-x}$Fe$_x$O$_3$ family of compounds.\cite{Troyanchuk2007, chiang2011, mihalik2016, mihalik2017}  In addition, 
the CMR material Ca$_{1-x}$Sm$_x$MnO$_3$ is a pseudo-perovskite manganite that showed phase separation into C-type and G-type 
antiferromagnetism.\cite{Mahendiran2000}  Given the different magnetic behavior above and below $x = 0.2$ for NdMn$_{1-x}$Fe$_x$O$_3$, 
the present neutron powder diffraction (NPD) work was undertaken to determine the magnetic structure of the mixed A-type and G-type magnetism 
on a three-dimensional cubic lattice NdMn$_{0.8}$Fe$_{0.2}$O$_3$, which is near the OAF phase approaching the tricritical doping. 

\begin{figure*}[t]
\begin{center}
\includegraphics[width=\linewidth]{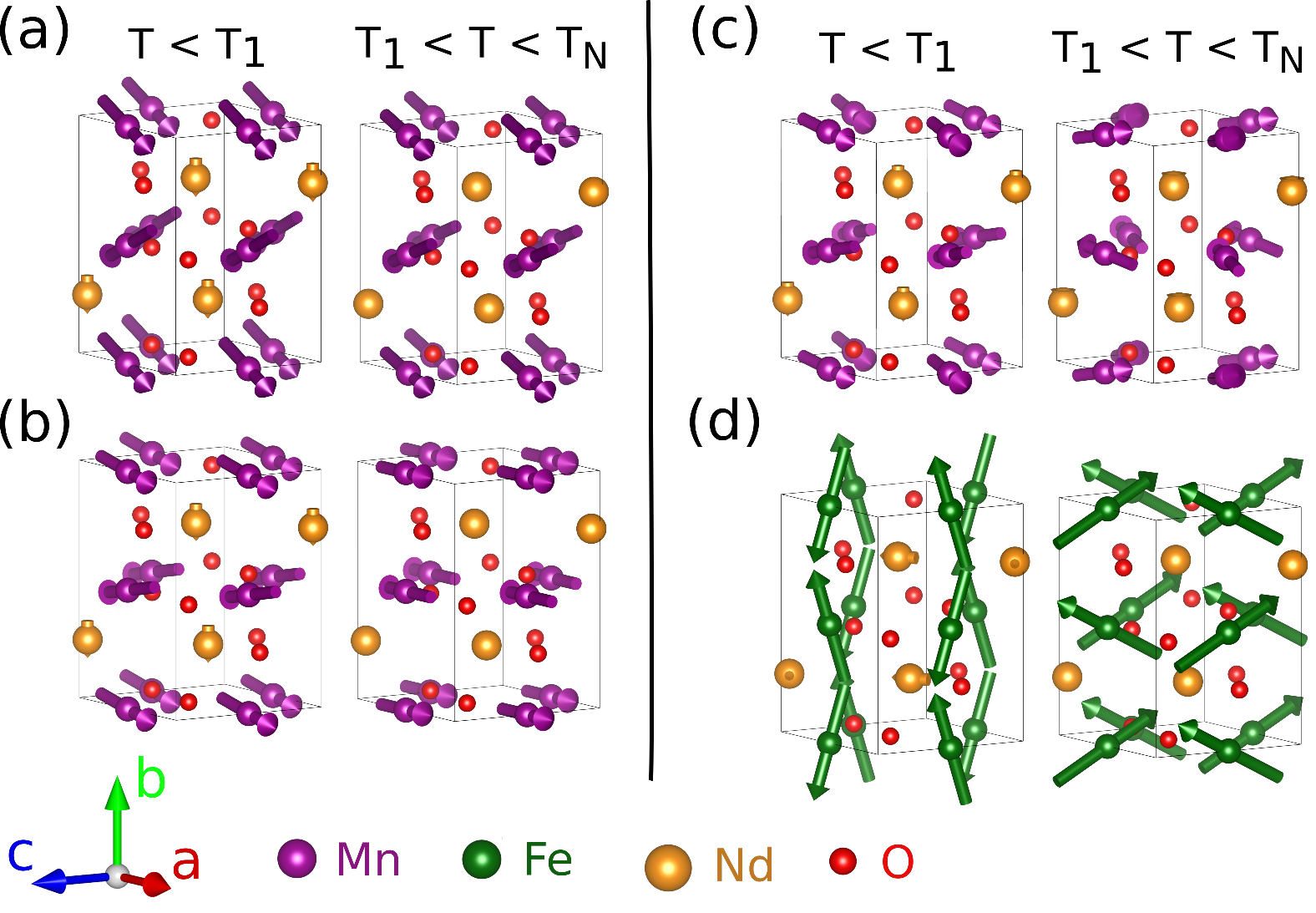}
\end{center}
\caption{\label{fig:1}  (Color online) Determined magnetic structures for (a)  NdMnO$_3$ ($T_1 \approx 13$~K; $T_{\mathrm N} \approx 78$~K) by Mu\~noz {\it et al.}\cite{Munoz2000}; (b) NdMnO$_3$ ($T_1 \approx 20$~K; $T_{\mathrm N} \approx 82$~K) by Chatterji {\it et al.}\cite{Chatterji2009}; (c) NdMn$_{0.8}$Fe$_{0.2}$O$_3$ ($T_1 \approx 13$~K; $T_{\mathrm N} \approx 59$~K)-- this work; (d) NdFeO$_3$ (left panel: $T_1 < 10$~K; right panel: $T_1 \approx 167$~K; $T_{\mathrm N} = 690$~K) from other authors.\cite{slawinski2005,bartolome1997,Przenioslo2006} The additional 2D projections of the magnetic structure are presented in Supplemental Material (SM) Fig. SM1. This drawing was made using the program VESTA.\cite{Vesta}}
\end{figure*}

Herein, our NPD data establish NdMn$_{0.8}$Fe$_{0.2}$O$_3$ orders magnetically with a magnetic propagation vector ${\bf k} = (000)$. The 3$d$ ions order into (A$_x$, F$_y$, G$_z$) magnetic structure and Nd$^{3+}$ ions order into (0, f$_y$, 0) magnetic structure in the interval 1.6~K~$< T < T_{\rm{N}} (\approx 59$~K), Fig.~\ref{fig:1}.  At $T_1 \approx 13$~K, a spin reorientation transition was observed, and this change is likely the origin of the anomalies reported from bulk measurements. \cite{Mihalik2013} The details of moment assignment will be discussed in the context of experimental and theoretical work and the presentation will be as follows, with the synthesis and experimental details presented in the next section, Section~\ref{experimental}. In Section~\ref{results}, results from temperature-dependent X-ray powder diffraction (XRPD) data 
are presented and have been used to study the structural transitions in the sample, while NPD has been used to extract the magnetic structure.  
Section~\ref{DFT} describes the energies from density functional theory (DFT) of four most-probable magnetic structures in the numerically tractable \emph{undoped}  
NdMnO$_3$.  Finally, a coherent picture of these results is discussed in Section~\ref{discussion} and summarized in Section~\ref{conclusion}.

\section{Experimental Methods, Analysis Protocols, and Computational Details \label{experimental}}

\subsection{Sample preparation and characterization}
Samples were prepared by a vertical floating zone (FZ) method in an optical mirror furnace. Starting materials consisted of  
high purity oxides of MnO$_2$ (purity 3N, Alpha Aesar), Nd$_2$O$_3$ (purity 3N, Sigma Aldrich), and Fe$_2$O$_3$ 
(purity 2N, Sigma Aldrich). These starting materials were mixed in a stoichiometric ratio, isostatically cold-pressed into rods, 
and subsequently sintered at $1100^{\circ}$C for 12 to 24 hours in air. The sintering procedure followed the solid state 
reaction preparation route,\cite{Pena2002} and the starting rods were already partially recrystallized after heat treatment. 
The floating zone experiment was performed using a 4-mirror optical furnace equipped with 1~kW halogen lamps and 
a pulling speed of 6~mm/h, a feeding speed of 4~mm/h, and a flowing (2~$\ell$/min) air atmosphere.   
The oxygen content was checked by iodometric titration, where a known amount of sample was dissolved in HCl solution (1:1, v/v) 
in the presence of KI. Immediately, the color of the solution turned to the yellow from I$_2$ arising from the oxidation of 
iodine anions.  Simultaneously with iodine oxidation, the manganese reduction proceeds to the 2+ oxidation state.  
Finally after completely dissolving the sample, the amount of iodine was determined by iodometric titration with Na$_2$S$_2$O$_3$ 
solution. Ultimately, a small excess of oxygen was detected, amounting to $\delta = 0.065$ for NdMn$_{0.8}$Fe$_{0.2}$O$_{3+\delta}$, 
and this excess oxygen indicates trace amounts of Mn$^{4+}$ ions are incorporated in our sample, which we will refer to as NdMn$_{0.8}$Fe$_{0.2}$O$_3$. 

It is generally accepted that solid solutions with a uniform chemical composition can be prepared by FZ techniques. 
In order to verify this assumption, the crystal structure of NdMn$_{0.8}$Fe$_{0.2}$O$_3$ was investigated by 
X-ray powder diffraction (XRPD), and all of the samples were established to be single-phase.  Next, 
two parts of the sample, one from the start and one from the end of the resulting ingot, were investigated by a 
scanning electron microscope (SEM), Mira III FE (produced by Tescan), which was equipped with an energy dispersive 
X-ray (EDX) analyzer, PentaFET Precision (produced by Oxford Instruments).  The SEM and EDX investigations revealed 
that both parts of the ingot were free of any inclusions, and no concentration gradient between the two ends of the crystal 
was detected.  Finally, the Nd:Mn:Fe ratio of 1:0.8:0.2, as determined by EDX analysis, was consistent with 
the other determinations within experimental uncertainties.

\subsection{Neutron diffraction studies}
The initial neutron powder diffraction (NPD) experiment was performed on the E6 neutron powder diffractometer at the Helmholtz-Zentrum 
Berlin (HZB). A freshly-ground powder sample with mass of about 5 g was enclosed, along with He exchange gas, in a vanadium container 
with a diameter of 5~mm. 
The settings of the diffractometer were as follows (downstream from the nuclear reactor): 30$^{\prime}$ Soller slit, 
pyrolytic graphite (PG) monochromator ($\lambda = 0.2454$~nm), PG filter, 30$^{\prime}$ Soller slit, 
sample enclosed in standard Orange cryostat, moving fan collimator, two position-sensitive detectors.  
Long scans were collected for $12.7^{\circ} < 2\theta < 102^{\circ}$ at temperatures of 1.6~K, 20~K, 35~K, and~65 K. 
In addition, several short scans $(12^{\circ} < 2\theta < 66^{\circ})$ were acquired in the temperature range 1.6~K $< T <$ 65~K.

An additional NPD experiment was performed on HB-3 triple-axis neutron spectrometer at the High Flux Isotope Reactor (HFIR) 
located at Oak Ridge National Laboratory (ORNL). During the experiment, the spectrometer was configured for elastic scattering 
with incident neutron energy 14.7~meV and $\lambda = 0.236$~nm.  Soller collimations of reactor $-$ 48$^{\prime}$ $-$ mono $-$ 
40$^{\prime}$ $-$ sample $-$ 40$^{\prime}$ $-$ analyzer $-$ 240$^{\prime}$ $-$ dectector were used, with slits 
optimized at a Bragg peak.  
For this experiment, approximately 20~g of freshly-ground powder 
along with He exchange gas was confined to an aluminum sample can, which was attached to a standard insert used with a 
standard Orange cryostat.  The one-day experiment focused on collecting data for $28^{\circ} \leq 2\theta \leq 33^{\circ}$ at 
temperatures of 1.7~K, 20~K, and 65~K.

\subsection{Low temperature XRPD studies} 
Low-temperature XRPD was performed using a refurbished Siemens D500 diffractometer equipped with a 
closed-cycle cryocooler (Sumitomo Heavy Industries) enabling measurements over a range of temperatures (3~K $\leq T \leq$ 300~K). 
Data were acquired with Cu-K$_{\alpha1,2}$ radiation and a Bragg-Brentano geometry with a source-sample and sample-detector 
distance of 330~mm. The sample environment consisted of a single crystalline sapphire sample holder, providing good thermal 
equilibration and low diffraction background, and He exchange gas to ensure homogeneous sample temperature. The measurements 
were performed in reflection geometry with a fixed divergence slit size, resulting in a primary beam with 0.44$^\circ$ divergence. 
A linear detector (MYTHEN 1K) along with an optimized integration procedure \cite{Kriegner2015} were used to avoid geometrical 
defocusing, while a Ni-foil was used to remove the K$_\beta$ radiation. 

\subsection{Diffraction analysis protocols and programs\label{protocols}}
All diffraction data were fitted using Le Bail and Rietveld methods implemented in the FullProf program.\cite{FullProf1993} The background was 
modeled by a polynomial function of maximum 5th order for room temperature (RT) XRPD data and NPD data. The background for the low temperature 
(LT) XRPD data was estimated manually due to a non-trivial shape caused by scattering by the windows of the sample chamber. 
Since the instrumental functions of the apparatuses were not established, the peak shape was modeled by a Thompson-Cox-Hastings 
pseudo-Voight function for the XRPD and by a Gaussian function for the NPD data. The initial conjectures of the profile functions for 
the NPD and LT XRPD data sets were obtained by Le Bail fits of a YIG standard and a LaB$_6$ NIST standard (standard number 660b), respectively. 
For describing the magnetic contributions to the NPD data, the standard magnetic form factors for Mn$^{3+}$, Fe$^{3+}$, and Nd$^{3+}$ ions 
that are incorporated in the FullProf program \cite{FullProf1993} were used.  All parameters allowed by the crystal symmetry of 
the crystallographic unit cell were refined. The symmetry analysis was performed using the program BasIreps, which is part of 
the FullProf Suite package of programs.\cite{FullProf2016} To find the global minimum of the best magnetic model we have generated large seeds of starting magnetic moments by home written java program. In these seeds, $\mu_{{\mathrm Mn}x}$, $\mu_{{\mathrm Mn}y}$ and $\mu_{{\mathrm Mn}z}$ starting values were tabulated within interval -4 $\mu_{\mathrm B}$ to 4 $\mu_{\mathrm B}$; $\mu_{{\mathrm Fe}x}$, $\mu_{{\mathrm Fe}y}$ and $\mu_{{\mathrm Fe}z}$ starting values within interval -5 $\mu_{\mathrm B}$ to 5 $\mu_{\mathrm B}$ and $\mu_{{\mathrm Nd}y}$ starting value within interval 0 -- 3.2 $\mu_{\mathrm B}$ with steps between 0.5 and 1.5 $\mu_B$, depending on the complexity of the calculations. Each point from this starting seed was then separately loaded into the FullProf program and refined for 10 cycles to get the representative values of R-factors.

\subsection{Computational details\label{abinitio}}
The first-principles (\emph{ab~initio}) calculations are based on the density functional theory\cite{Kohn64} within the 
single-electron framework and are used herein to treat the pure stochiometric NdMnO$_3$ compound. 
The VASP (Vienna Ab-initio Simulation Package) package,\cite{Kresse96a, Kresse96b}   
a plane-wave pseudopotential code, was used to perform spin-polarized calculations including the spin-orbit 
interaction. Projector-augmented-wave pseudopotentials were used for Nd, Mn, and O atoms with the electronic 
valence configurations of [Xe] $4f^3$ (oxidation state 3+), [Ar] $3d^5$ $4s^2$, and [He] $2s^2$ $2^4$, respectively. 
General gradient-corrected exchange-correlation functionals parametrized by Perdew-Burke-Ernzerhof 
(PBE)\cite{Perdew96} and a plane-wave cut-off of 600~eV were employed. The unit cell was sampled with a $k$-point mesh of 
$6\times 4\times 6$ generated according to the scheme proposed by Monkhorst and Pack.\cite{Monkhorst}  The convergence criteria 
for the total energies and forces were set to 10$^{−6}$~eV and 10$^{−4}$~eV/\AA, respectively. Electron 
correlation beyond the PBE was taken into account within the framework of so-called GGA +U method and the 
approach proposed by Dudarev \emph{et al.}\cite{Dudarev98} Calculations were carried out with the Coulomb 
repulsion $U$ and the exchange parameter $J$ in the range of $0.1-10$~eV for the {\it d}- and {\it f}-electrons of Mn and Nd atoms. 
The spin-orbit interaction of the valence states was taken into account.  
%
%%%%%%%%%%%%%%%%%%%%%%%%%%%%%%%%%%%%%%%%%%%%%%%%%%%%%%%%%%%%%%%%%%%%%
%
\section{Results \label{results}}

\subsection{Crystal structure refinement\label{crystal-structure}}
The crystal structure of NdMn$_{0.8}$Fe$_{0.2}$O$_3$ was refined from the XRPD and NPD data at room temperature. 
The process was done by first treating the XRPD and the NPD data sets separately, and subsequently, these two diffractograms were co-refined. 
Since NdMnO$_3$ and NdFeO$_3$ adopt the same crystal structure\cite{Munoz2000, slawinski2005} (orthorhombic structure, space group $Pnma$,  
with atomic positions: Mn/Fe: 4$b$; Nd: 4$c$; O$_1$: 4$c$; O$_2$: 8$d$), the crystal structure of NdMnO$_3$ reported by Mu\~{n}oz {\it et al.} \cite{Munoz2000} 
was used as a starting model. The Rietveld fit using this model resulted in low $R$-factors and to the crystallographic 
parameters presented in Table~\ref{table1}, and these results indicate that NdMn$_{0.8}$Fe$_{0.2}$O$_3$ maintains the 
structure of the parent compound NdMnO$_3$. To determine the rare-earth deficiency,\cite{Cherepanov1995} 
the occupancy factor of the Nd atoms was allowed to vary during the first stages of the refinement 
of the NPD data, and the value converged to 1.04(1), thereby indicating no appreciable evidence of Nd non-stoichiometry. 
Consequently, the Nd site was considered to be fully occupied in the next stages of refinement and for the processing of 
all experimental data collected below room temperature.

\begin{table}[b!]
\begin{center}
\begin{tabular}{cccc}
\hline
& XRPD&	NPD&	co-refined fit\\
\hline
a (nm)&	0.5772(2)&	0.5781(4)&	0.5771(1)\\
b (nm)&	0.7600(3)&	0.7630(6)&	0.7604(2)\\
c (nm)&	0.5419(2)&	0.5439(4)&	0.5422(2)\\
V (nm$^3$)&	0.2377(2)&	0.2399(3)&	0.2379(4)\\
\hline
x$_{\rm Nd}$&	0.065(2)&	0.064(6)&	0.058(5)\\
z$_{\rm Nd}$&	0.984(3)&	0.973(1)&	0.975(4)\\
x$_{\rm O1}$&	0.487(9)&	0.478(3)&	0.481(3)\\
z$_{\rm O1}$&	0.081(8)&	0.087(9)&	0.095(5)\\
x$_{\rm O2}$&	0.326(7)&	0.315(5)&	0.316(4)\\
y$_{\rm O2}$&	0.038(7)&	0.040(6)&	0.027(7)\\
z$_{\rm O2}$&	0.705(7)&	0.708(6)&	0.705(4)\\
\hline
\end{tabular}
\caption{The comparison of the crystallographic parameters of NdMn$_{0.8}$Fe$_{0.2}$O$_3$ as obtained from the 
different diffraction techniques and at room temperature. The resulting $R$-factors are: $R_p$ = 20.0, 
$R_{wp}$ = 25.8, $R_{exp}$ = 14.3, $\chi^2$ = 3.24 for XRPD data; $R_p$ = 4.81, $R_{wp}$ = 6.73, $R_{exp}$ = 4.23, 
$\chi^2$ = 2.53 for NPD data; and $R_p$ = 20.4, $R_{wp}$ = 19.6, $R_{exp}$ = 1.97, $\chi^2$ = 4.10 for co-refined fit.
\label{table1}}
\end{center}
\end{table}

\begin{figure}[t!]
\begin{center}
\includegraphics[width=0.95\linewidth]{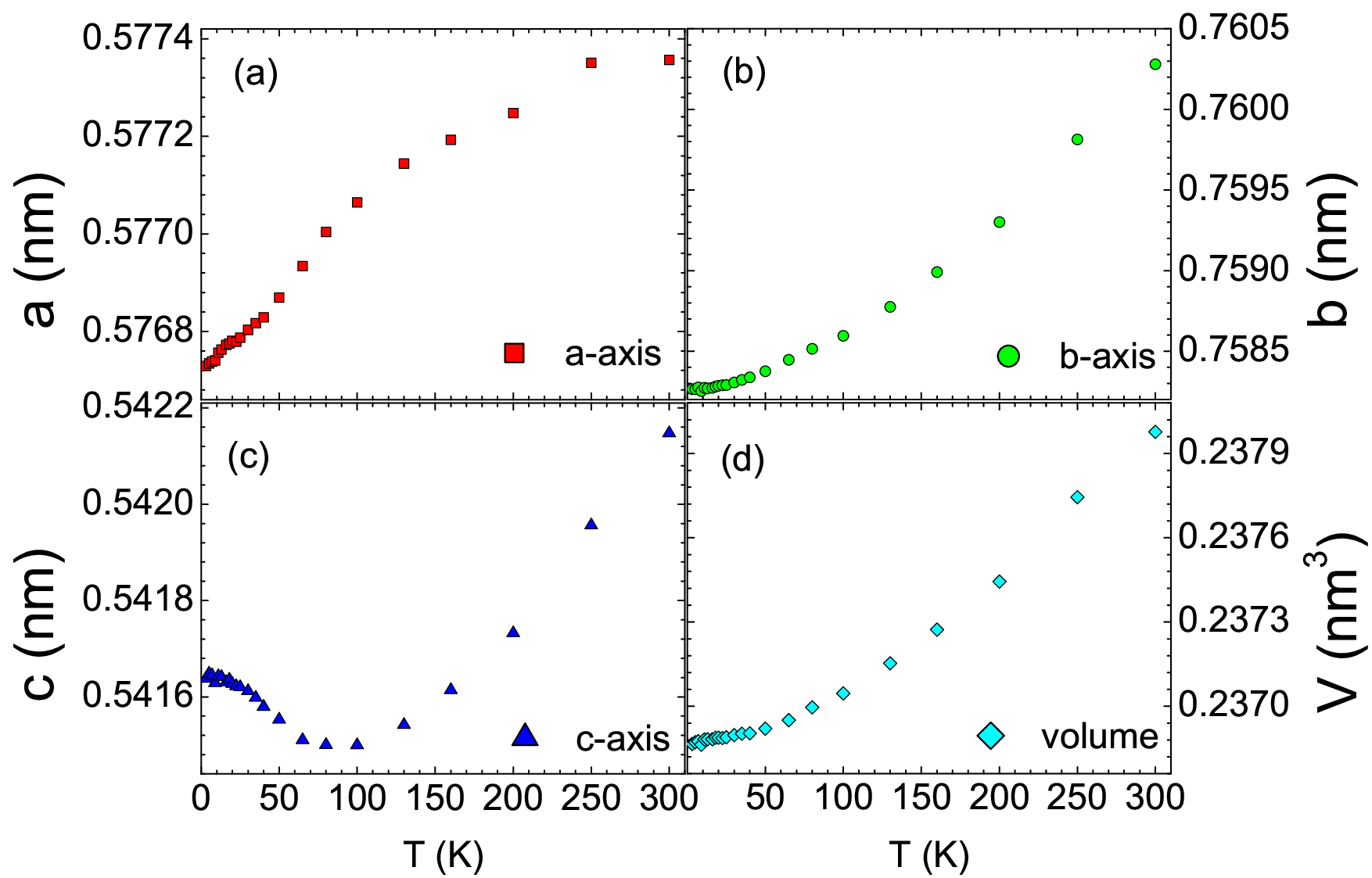}
\caption{(Color online) The temperature evolution of the crystallographic parameters as determined by fitting the low temperature 
XRPD data using the Le Bail method. Raw data used for the analysis are presented in supplementary Fig.~SM2.
\label{Fig:2}}
\end{center}
\end{figure}

The calculated lattice parameters decrease monotonically in the temperature range 80~K $\leq T \leq$ 300~K (Fig.~\ref{Fig:2}), 
and the observed changes are consistent with thermal contraction. Below 65(10)~K, there is a clear increase of the 
$c$-axis length, while the $a$-axis, $b$-axis, and volume changes are more subtle.  In comparison, bulk probes found $T_{\mathrm N} = 58.6(5)$~K.\cite{Mihalik2013} The similar temperature evolution of the crystallographic parameters was observed in the case of NdFeO$_3$ compound in the spin reorientation region,\cite{slawinski2005} but in the case of NdMnO$_3$, the sudden drop of all three crystallographic parameters was observed at T$_{\mathrm N}$.\cite{Chatterji2009} Therefore, the magnetoelastic coupling in NdMn$_{0.8}$Fe$_{0.2}$O$_{3}$ is different from NdMnO$_3$, but can be similar to NdFeO$_3$. No extra peaks were observed at temperatures below 300~K [see Fig.~SM2], and  no essential shifts of fractional coordinates (see Fig.~SM3), which would imply the presence of  
spin-rotation/octahedral-rocking that was detected in NdFeO$_3$,\cite{Belov1972,Koshizuka1988} were observed. 
These results imply that no structural phase transitions exist in the temperature range 3~K $\leq T \leq 300$~K. 
Consequently, when determining the magnetic structure of NdMn$_{0.8}$Fe$_{0.2}$O$_3$ (see next section), 
the crystal structure was fixed to be the orthorhombic structure, space group $Pnma$. 

\subsection{Magnetic structure refinement}
\begin{figure}[t]
\begin{center}
%\hspace{-4.25cm}
\includegraphics[width=0.95\linewidth]{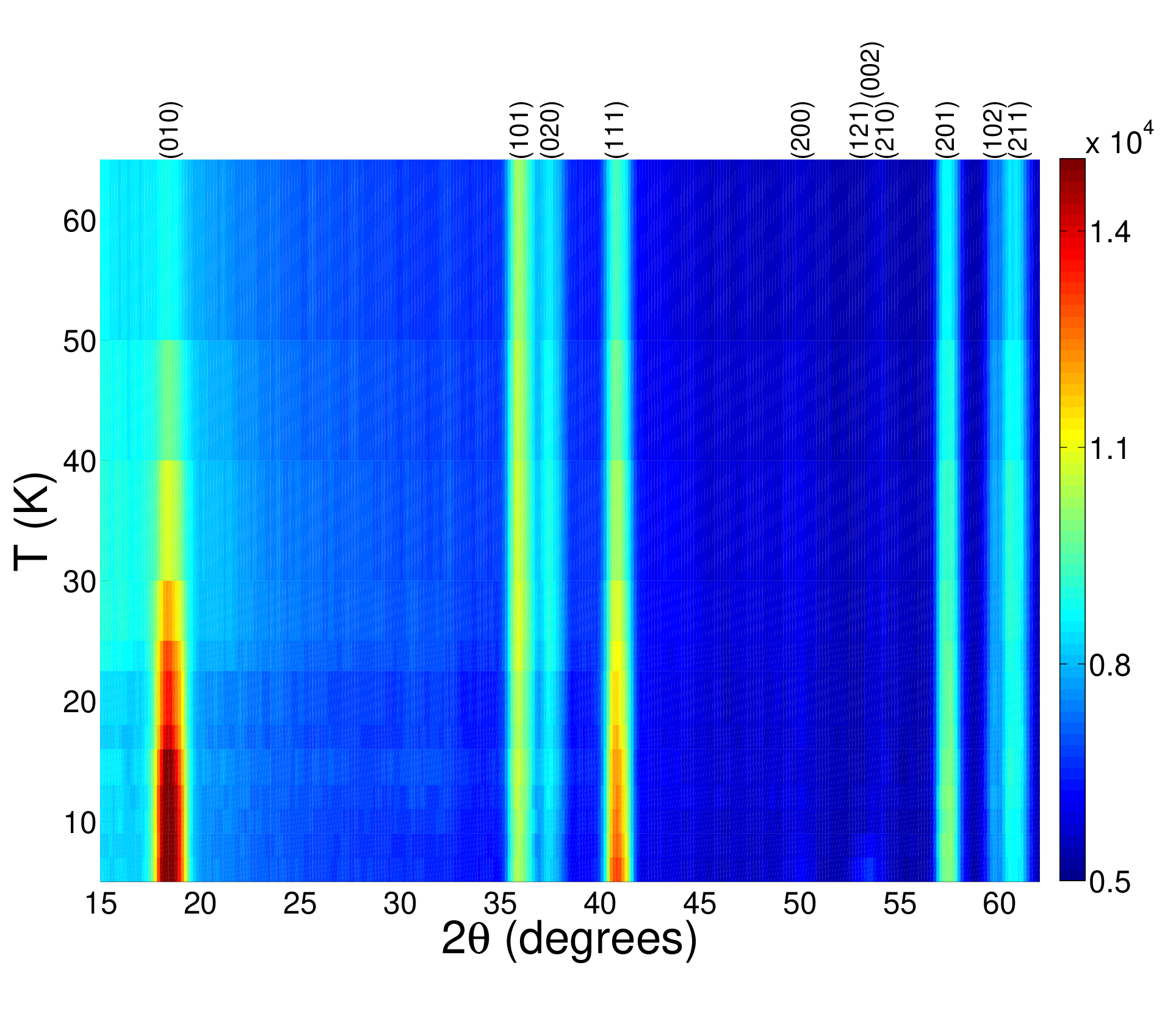}
\end{center}
\caption{\label{fig:3}  (Color online) Temperature variation of the diffraction intensities as a function of $2\theta$ for 
NdMn$_{0.8}$Fe$_{0.2}$O$_{3}$. The color scale for the observed intensity is given to the right of the main plot 
of data collected on the E6 diffractometer at HZB.}
\end{figure}

\begin{figure}[t]
\begin{center}
\includegraphics[width=0.95\linewidth]{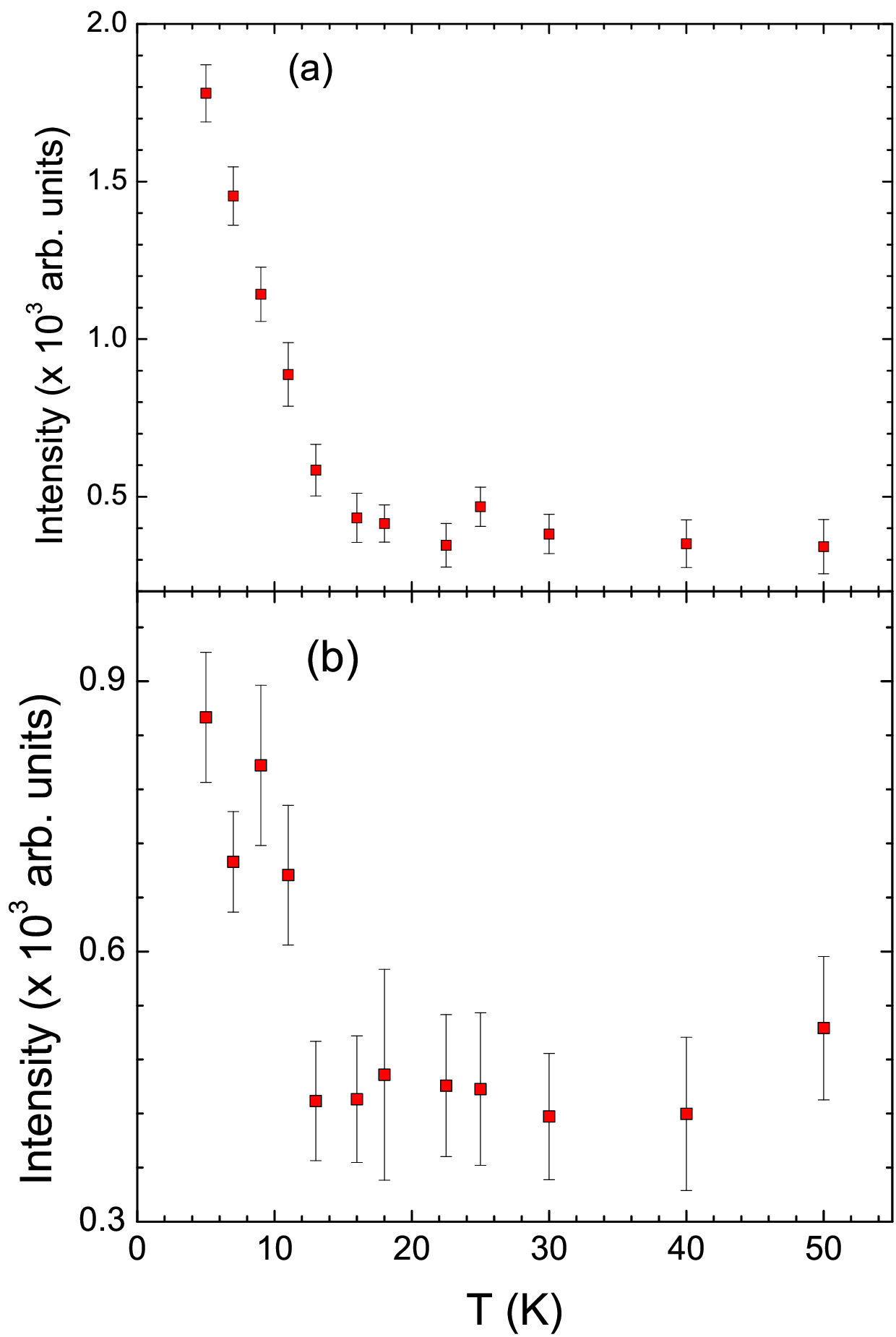}
\end{center}
\caption{\label{fig:4}  (Color online) Temperature variation of the integrated intensities of the (a) overlapping (121), (002), and (210) reflections (b) (200) reflection. Data collected on the E6 diffractometer at HZB.}
\end{figure}

Although the XRPD study below 65~K suggests the orthorhombic symmetry of the crystal structure remains unchanged, 
the NPD experiment revealed that intensities of some reflections increase with decreasing temperature, for example the (111) 
reflection, and a gradual increase of intensity appears, for example on the (010) reflection, which is forbidden by the 
space group $Pnma$\cite{ITC}  (Fig.~\ref{fig:3}). These changes are associated with magnetic ordering setting in below 
$T_{\mathrm N} = 58.6(5)$~K, which is in agreement with our AC susceptibility and magnetization measurements.\cite{Mihalik2013}  
Since all magnetic reflections can be indexed by integer $hkl$ indices, the magnetic ordering wavevector is ${\bf k} = (000)$.  
Furthermore, below $T_1 \approx 13$~K, remarkable changes in the intensity of some magnetic peaks, namely the overlapping 
(121), (002), and (210) reflections and the (200) reflection are observed, see Fig.~\ref{fig:4}.  
These increases of intensities indicate the magnetic structure is evolving and/or the other magnetic ion is ordering. 
These low temperature changes of the diffraction pattern onset with an anomaly detected in the AC susceptibility at 
$T_1 = 13$~K.\cite{Mihalik2013}  For $T < T_1$, no additional magnetic peaks appear, and the magnetic structure is described 
by the same propagation vector ${\bf k} = (000)$.

Assuming no spin-lattice induced change in the space group, the possible magnetic modes compatible with the 
crystal symmetry have been obtained using the program BasIreps.\cite{FullProf2016}  
For ${\bf k} = (000)$, the little group, $\Gamma_{\mathrm{k}}$, coincides with the space group $Pnma$. Of the eight 
$\Gamma_i$'s for the $Pnma$ 4$b$ position of the Mn and Fe, four allow magnetic order such that
\begin{equation}
\Gamma_{\mathrm{Mn/Fe}}\;=\;3\,(\Gamma_1 + \Gamma_3 + \Gamma_5 + \Gamma_7)\;\,.
\end{equation}
For Nd atoms on the $4c$ site, the decomposition is
\begin{equation}
\Gamma_{\mathrm{Nd}}\;=\; \Gamma_1 + \Gamma_4 + \Gamma_5 + \Gamma_8 + 2\,(\Gamma_2 + \Gamma_3 + \Gamma_6 + \Gamma_7)\;\,.
\end{equation}
The basis vectors obtained for each irreducible representation $\Gamma_i$ are reported in the Appendix, see Table~\ref{magnSpaceGroup}.

Based on the results for LaMnO$_3$\cite{Moussa1996} and extapolated generically to $R$MnO$_3$, it is widely accepted that the 
Mn-sublattice orders at much higher temperatures than those where the $R$ ions become polarized due to the $R$--Mn interaction. 
However, several different magnetic structures for NdMnO$_3$ have been reported by various 
groups, including the possibility that Nd ions order already at T$_N$.\cite{Munoz2000,Chatterji2009,Jandl2003,Bartolome2005}  For this reason, all possible magnetic structures allowed by 
the basic symmetry constraints were considered, including the independent ordering of Mn/Fe- and Nd-sublattices and the 
plausible case that the Nd moments remain disordered (denoted as $\Gamma_0$ state). In total, 36~model structures were compared with the NPD data sets 
collected at $T$ = 1.6 K at HZB.  When all experimentally detected peaks were described by a model structure and no extra peaks with 
intensities higher than the experimental noise were generated, then plausible matches were considered to be established
between a model structure and the data.  The next step involved Rietveld analysis starting with each plausible model structure.  
The results of this comprehensive analysis are summarized and 
tabulated in Table~SM1, where the magnetically ordered state notation, 
$\Gamma_{i\mathrm{Mn/Fe}}\Gamma_{j\mathrm{Nd}}$ is defined and cross-referenced.  
This analysis resulted in four magnetic structures whose refined R-factors did not distinguish any 
single structure as the unambiguous solution. These four magnetic structures are: $\Gamma_{5\mathrm{Mn/Fe}}\Gamma_{7\mathrm{Nd}}$, $\Gamma_{5\mathrm{Mn/Fe}}\Gamma_{3\mathrm{Nd}}$, $\Gamma_{5\mathrm{Mn/Fe}}\Gamma_{5\mathrm{Nd}}$ and $\Gamma_{5\mathrm{Mn/Fe}}\Gamma_{0\mathrm{Nd}}$.

In all cases, the best fit for $T = 1.6$~K is found to be $\Gamma_{\rm{5Mn/Fe}}$ for Mn/Fe sublattice, but the goodness of fit parameters cannot unambiguously distinguish between $\Gamma_{\rm{0Nd}}$, $\Gamma_{\rm{3Nd}}$, $\Gamma_{\rm{5Nd}}$, and $\Gamma_{\rm{7Nd}}$. The absence of any additional structural phase transitions in the XRPD is suggestive that the magnetic space group does not change at $T_1$. Also extrapolating from LaMnO$_3$,\cite{Moussa1996} it is generally accepted the Mn sublattice orders at $T_{\mathrm N}$. This inference implies that if the Nd  sublattice order, then it should order within the same magnetic space group as Mn sublattice. Consequently, the $\Gamma_{\rm{3Nd}}$ and $\Gamma_{\rm{7Nd}}$ are not physically allowed, but $\Gamma_{\rm{0Nd}}$, and $\Gamma_{\rm{5Nd}}$ remain as plausible configurations. Note that the $\Gamma_{\rm{0Nd}}$ notation means the Nd ions do not order into long range magnetic structure, which is consistent with the statement that at $T_1$ the magnetic structure evolves, but no additional ion orders at that temperature. Therefore, the magnetic space group of NdMn$_{0.8}$Fe$_{0.2}$O$_3$ is assigned to be $Pn^{\prime}ma^{\prime}$.

Considering the 4$b$ transition metal site, the $\Gamma_5$ representation can host A-type (as for NdMnO$_3$) 
and G-type (as for NdFeO$_3$) antiferromagnetism. Since 4$b$ site hosts Mn and Fe ions, we have tried to fit independently Mn and Fe magnetic moments. Second analysis was done with Mn magnetic moments constrained to the (A$_x$, F$_y$, 0) magnetic structure (as for NdMnO$_3$) and the Fe magnetic moments to (0, F$_y$, G$_z$) magnetic structure (as for NdFeO$_3$). Despite the fact that large seeds of initial fitting parameters were used (see section \ref{protocols}), \emph{all} fits in both cases converged to unphysical results. Consequently, these two options were rejected, leaving the only two possibilities that either the Mn or the Fe ions exclusively order. Since $T_{\mathrm N}$ for $x$ = 0.2 is smaller than for $x$ = 0 and a minimum of $T_{\mathrm N}$ is expected at concentrations $x \geq$ 0.25,\cite{Mihalik2013}  one can expect that the Fe ions act only as a perturbation and the magnetism is mainly driven by the Mn ions. As a result on the $4b$ site, only the Mn ions order and the possible magnetic ordering can be $\Gamma_{5\mathrm{Mn}}\Gamma_{0\mathrm{Nd}}$, or $\Gamma_{5\mathrm{Mn}}\Gamma_{5\mathrm{Nd}}$. Finally, the large seed initial fitting parameters test (see Section \ref{protocols}) for structures $\Gamma_{5\mathrm{Mn}}\Gamma_{0\mathrm{Nd}}$ and $\Gamma_{5\mathrm{Mn}}\Gamma_{5\mathrm{Nd}}$ revealed 6 local minima in the entire parameter space, where the fitting parameters resulted in physically meaningful values (see Table SM2, where the numbering of the minima is also defined). From these 6 candidates, only two plausible descriptions emerge, see Table~SM2 for details.

\subsection{Temperature dependences of the magnetic moments}

\begin{figure}[t!]
\begin{center}
\includegraphics[width=\linewidth]{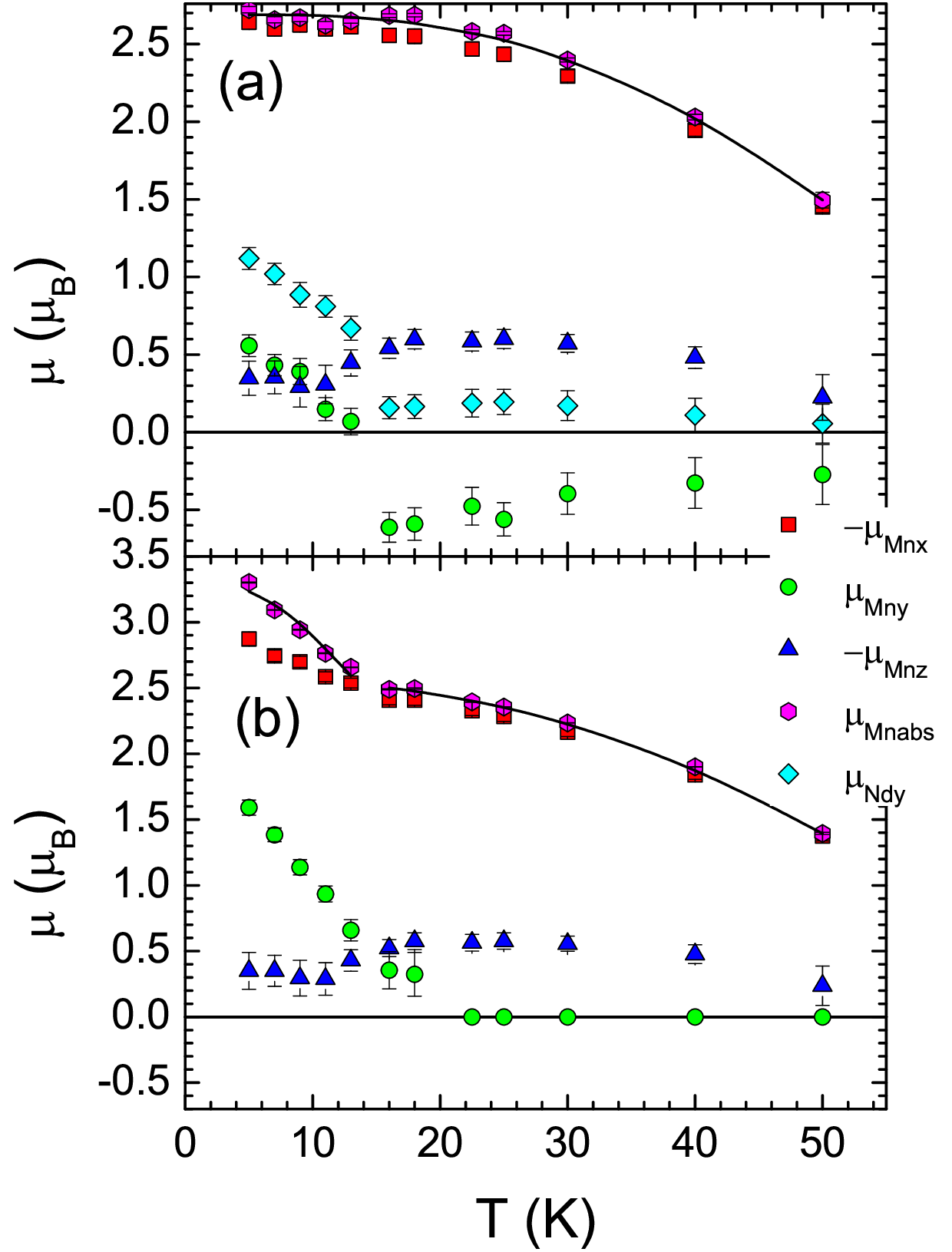}
\end{center}
\caption{\label{fig:5}  (Color online) The temperature evolution of the magnetic components and absolute value of the Mn  
magnetic moment for the magnetic structure (a) $\Gamma_{5\mathrm{Mn}}\Gamma_{5\mathrm{Nd}}$; (b) $\Gamma_{5\mathrm{Mn}}\Gamma_{0\mathrm{Nd}}$ The solid lines represent the best fits according 
to Eq.~\ref{brillouin}.}%, and the dashed lines are an extrapolation from the fitting region to lower temperatures.}
\end{figure}

The temperature dependences of the two remaining candidates for the magnetic structure are shown in Fig.~\ref{fig:5}. In the case of $\Gamma_{5\mathrm{Mn}}\Gamma_{5\mathrm{Nd}}$, the Nd ions order at T$_{\mathrm N}$, but $\mu_{\mathrm{Nd}y}$ exhibits an abrupt increase at $T_1$ and $\mu_{\mathrm{Mn}y}$ flips to the opposite direction at the same temperature. In case of $\Gamma_{5\mathrm{Mn}}\Gamma_{0\mathrm{Nd}}$, $T_1$ can be attributed to the evolution of the $\mu_{\mathrm{Mn}y}$ component. 

\begin{figure}[t!]
\begin{center}
%\hspace{-5.5cm}
\includegraphics[width=\linewidth]{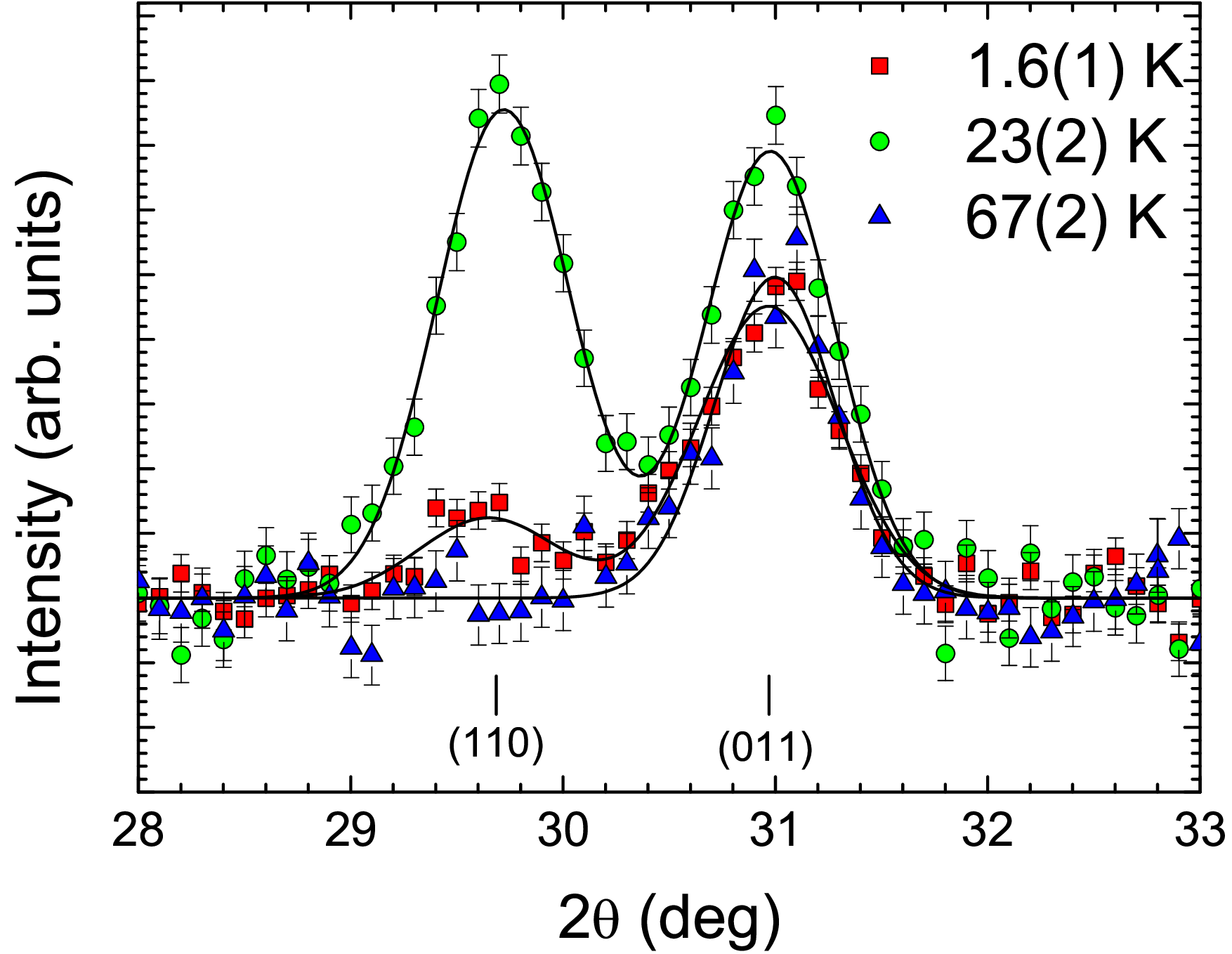}
\end{center}
\caption{\label{fig:6}  (Color online) The NPD patterns obtained, after 
subtracting a linear background, in the vicinity of the (110) and (011) reflections 
as measured on the HB-3 triple-axis instrument at ORNL. The lines represent Gaussian fits of the intensity.}
\end{figure}

In both cases, $T_1$ is also connected with the continuous decrease of the $\mu_{\mathrm{Mn}z}$ component, which is typical for a spin-reorientation phase transition. Since this effect is rather weak, additional tests are needed to distinguish if the effect is real. The permitted reflections for G$_z$ mode have constraints $k$ is odd and $h + l = 2n + 1$. Consequently, the strongest contribution to the magnetic signal from the G$_z$ mode should be observed for the (110) reflection.  Since the intensity of the (110) reflection is close to the background of NPD patterns collected at HZB, an additional NPD experiment focused on resolving this issue was performed at ORNL.
The data from this experiment unambiguously show that the magnetic signal on the (110) reflection, Fig.~\ref{fig:6}, is stronger at 23(2)~K than at 1.6(1) K, thereby confirming that the spin reorientation phase transition is real effect.

%continue here

According to a molecular field model,\cite{Kozlenko2004}  
the temperature evolution of the total magnetic moment $\mu(T)$ follows a self-consistent expression written as

\begin{equation}
\mu(T)\;=\;\mu_0 B_J\left( \,\frac{3J}{(J+1)}\; \frac{\mu(T)}{\mu_0} \; \frac{T_{\mathrm{N}}}{T}\,\right)\;\;\;, \label{brillouin}
\end{equation}
where $\mu_0$ is the magnetic moment at $T = 0$, $B_J$ is the Brillouin function, $T_{\mathrm{N}}$ is the ordering temperature.  Fits according to Eq. \ref{brillouin} yeld $T_{\rm{N}}$ = 58.7~K and $T_{\rm{N}}$ = 57.6~K for the $\Gamma_{5\mathrm{Mn}}\Gamma_{5\mathrm{Nd}}$ and $\Gamma_{5\mathrm{Mn}}\Gamma_{0\mathrm{Nd}}$ magnetic structures, respectively. The value of $T_{\rm{N}}$ for the $\Gamma_{5\mathrm{Mn}}\Gamma_{5\mathrm{Nd}}$ magnetic structure is closer to $T_{\rm{N}} = 58.6(5)$~K, obtained from bulk magnetization measurements.\cite{Mihalik2013} The magnetic moment of the Mn sublattice extrapolated to $T = 0$~K is 2.69 $\mu_{\rm B}$ for $\Gamma_{5\mathrm{Mn}}\Gamma_{5\mathrm{Nd}}$, and this result is lower than $\mu_{\rm Mn}$ = 3.87(3) $\mu_{\rm B}$ reported for LaMnO$_3$.\cite{Moussa1996} On the other hand for the $\Gamma_{5\mathrm{Mn}}\Gamma_{0\mathrm{Nd}}$ structure, the magnetic moment in the $T \rightarrow 0$ limit is $\mu_{\rm Mn}$ = 3.27 $\mu_{\rm B}$, which is much closer to $\mu_{\rm Mn}$ reported for LaMnO$_3$.\cite{Moussa1996} However, the data for the $\Gamma_{5\mathrm{Mn}}\Gamma_{0\mathrm{Nd}}$ structure are not well-modeled by a single Brillouin function, and this result may suggest the presence of a phase transition at $T_1$. Since specific heat data of the NdMn$_{0.8}$Fe$_{0.2}$O$_3$ compound show no anomaly at $T_1$,\cite{Mihalik2013} it is plausible  that $T_1$ is not connected with a phase transition. Additionally, the R-factors determined at 20~K are much lower for the $\Gamma_{5\mathrm{Mn}}\Gamma_{5\mathrm{Nd}}$ magnetic structure than for the $\Gamma_{5\mathrm{Mn}}\Gamma_{0\mathrm{Nd}}$ (see Table SM3). Finally, a recent backscattering experiment\cite{daniel2018} resolved a non-zero polarization of the Nd ions for $T < T_{\mathrm{N}}$, so the $\Gamma_{5\mathrm{Mn}}\Gamma_{0\mathrm{Nd}}$ magnetic configuration can be eliminated as a physical option, thereby leaving $\Gamma_{5\mathrm{Mn}}\Gamma_{5\mathrm{Nd}}$ as the only possible description.  Specifically, the magnetic configuration is (A$_x$, F$_y$, G$_z$) for Mn ions and (0, f$_y$, 0) for Nd sublattice in the whole temperature range 1.6~K~$\leq T \leq T_{\mathrm N}$.  
%
%%%%%%%%%%%%%%%continue here%%%%%%%%%%%%%%%%%%%%%%%%%%%%%%%%%%%%%%%%
%

\section{N\lowercase{d}M\lowercase{n}O$_3$ Magnetic structure by density funtional theory calculations \label{DFT}}
In the previous section, the magnetic structure of NdMn$_{0.8}$Fe$_{0.2}$O$_3$ was experimentally established to be 
$Pn^{\prime}ma^{\prime}$ (A$_x$,F$_y$,G$_z$)~+~(0, f$_y$, 0). Since our analysis of NdMn$_{0.8}$Fe$_{0.2}$O$_3$ resulted in a magnetic structure different than the most-accepted magnetic structure of NdMnO$_3$ phase, but there have been some inconsistencies in the literature about the NdMnO$_3$ magnetic structure,\cite{Munoz2000,Chatterji2009,Jandl2003,Bartolome2005} this section describes 
a theoretical approach to understand the magnetic structure of the pure NdMnO$_3$. 

To start, varying $U$ and $J$ values as initial parameters yielded $U_{\mathrm{Nd}}= 5$~eV, 
$J_{\mathrm{Nd}}=0.1$~eV for the $f$-shell of Nd, and with $U_{\mathrm{Mn}}= 10$~eV and 
$J_{\mathrm{Mn}}=2$~eV for the $d$-shell of Mn atoms to preserve insulating behavior 
and the magnetic moment length.  These calculations ultimately led to values for the magnetic moments 
$\mu_{\mathrm{Nd}} = 1.4~\mu_{\mathrm{B}}$ and $\mu_{\mathrm{Mn}} = 3.9~\mu_{\mathrm{B}}$ and to the band gap of 1.75~eV, 
and these results are comparable to the experimental observations.\cite{Mihalik2013,Munoz2000,Chatterji2009,Shetkar2010}    
These first-principle calculations revealed that the total magnetic moment of the 
Nd atom is reduced by the large orbital moment \emph{ca.}~1.5~$\mu_{\mathrm{B}}/$atom 
that is antiparallel with respect to its spin moment of 2.9~$\mu_{\mathrm{B}}/$atom.

Next, the crystallographic structure was optimized by performing a complete relaxation of the 
lattice vectors as well as the atomic positions and internal degrees of freedom. During this initial optimization, 
two different types of exchange interactions, either antiferromagnetic or ferromagnetic between Mn 
spins, were considered. This crystallographic optimization led, in both cases, to a decrease of the space group symmetry 
from $Pnma$ to $P2_1/c$. However, a closer look to the optimized structure revealed, in both cases, 
only the minor shifts of the $y$-position of the Nd ion, from 1/4 to $0.250\epsilon$, where $\epsilon$ stands for non-zero digit 
lower than 5, but such small shifts are below the precision of the experimental methods.  
Consequently, the orthorhombic symmetry (space group $Pnma$) was employed in all of the 
following steps of the calculations.
 
Ultimately, two different crystallographic structures were obtained, and for the purposes of these 
numerical studies, these structures are denoted as p$_{\rm{AF}}$ and p$_{\rm{F}}$.  
When assuming antiferromagnetic interactions between the Mn ions, the p$_{\rm{AF}}$ state is identified with lattice parameters 
$a = 0.5968$~nm, $b = 0.7702$~nm, and $c = 0.5500$~nm.  Conversely, when assuming ferromagnetic interactions between 
the Mn ions, the p$_{\rm{F}}$ configuration is found with lattice parameters 
$a = 0.5987$~nm, $b = 0.7659$~nm, and $c = 0.5505$~nm. Both 
structures were obtained by relaxing all degrees of freedom, while only the initial magnetic pattern was different. 
The optimized lattice parameters are roughly 1.5\% to 2.5\% higher than the experimentally determined lattice parameters 
as presented in Section~\ref{crystal-structure} and Table~\ref{table1}, and p$_{\rm{exp}}$ will designate the observed lattice. 
These results are consistent with the well-known over-binding effects of the GGA (PBE) exchange-correlation 
approximation employed for this work, see Section~\ref{abinitio} for details. Furthermore, such small differences 
in lattice parameters indicate a very good match between theory and experiment. 

Finally, four magnetically ordered states, namely 
$\Gamma_{\mathrm{5Mn}}\Gamma_{\mathrm{7Nd}}$ [Mn (A$_x$,~F$_y$,~G$_z$) and Nd (f$_x$,~0,~0)], 
$\Gamma_{\mathrm{5Mn}}\Gamma_{\mathrm{3Nd}}$ [Mn (A$_x$,~F$_y$,~G$_z$) and Nd (0,~0,~f$_z$)], 
$\Gamma_{\mathrm{5Mn}}\Gamma_{\mathrm{5Nd}}$ [Mn (A$_x$,~F$_y$,~0) and Nd (0,~f$_y$,~0)], and 
$\Gamma_{\mathrm{5Mn}}\Gamma_{0{\mathrm{Nd}}}$ [Mn (A$_x$,~F$_y$,~G$_z$) and with Nd disordered],
were introduced for p$_{\rm{AF}}$, p$_{\rm{F}}$, and p$_{\rm{exp}}$ structures and only the electronic degrees of freedom were converged, 
{\it i.e.}~the atomic positions were kept fixed. 

These magnetic structures probe the direction of the Nd polarization, and antiferromagnetic modes that are allowed within 
the specific irreducible representations are suppressed for this analysis.  

The calculated total energies for each crystallographic structure, 
p$_{\rm{AF}}$, p$_{\rm{F}}$, and p$_{\rm{exp}}$, and the plausible magnetically ordered states are summarized in Table~\ref{magorder4}. 
From this tabulation, one immediately notices the total energies of all three structures 
p$_{\rm{AF}}$, p$_{\rm{F}}$, and p$_{\rm{exp}}$ with magnetic ordering of $\Gamma_{\mathrm{5Mn}}\Gamma_{\mathrm{7Nd}}$ and 
$\Gamma_{\mathrm{5Mn}}\Gamma_{\mathrm{3Nd}}$ are higher in energy than other two structures 
($\Gamma_{\mathrm{5Mn}}\Gamma_{\mathrm{5Nd}}$ and $\Gamma_{\mathrm{5Mn}}\Gamma_{0{\mathrm{Nd}}}$), 
with exception p$_{\rm{AF}}$ and $\Gamma_{\mathrm{5Mn}}\Gamma_{\mathrm{5Nd}}$.  Consequently, these 
higher energy results are excluded from further consideration.  It is important to note that the 
p$_{\rm{AF}}$ and $\Gamma_{\mathrm{5Mn}}\Gamma_{0{\mathrm{Nd}}}$ structure may have a higher energy  
because of the lack of suitable charge convergence with respect to the other entries in the table. 

\begin{table}[b!]
\begin{center}
\begin{tabular}{cccccc}
\hline \hline
magnetic state & Mn & Nd & p$_{\rm{AF}}$ & p$_{\rm{F}}$ & p$_{\rm{exp}}$ \\
\hline
$\Gamma_{\mathrm{5Mn}}\Gamma_{\mathrm{7Nd}}$& (A$_x$,~F$_y$,~G$_z$) & (f$_x$,~0,~0) & 10.2 & 12.5 & 9.5 \\
$\Gamma_{\mathrm{5Mn}}\Gamma_{\mathrm{3Nd}}$& (A$_x$,~F$_y$,~G$_z$) & (0,~0,~f$_z$) & 25.5 & 41.2 & 10.7 \\
$\Gamma_{\mathrm{5Mn}}\Gamma_{\mathrm{5Nd}}$& (A$_x$,~F$_y$,~0) & (0,~f$_y$,~0) & 23.5 & 0.0 & 0.0 \\
$\Gamma_{\mathrm{5Mn}}\Gamma_{0{\mathrm{Nd}}}$& (A$_x$,~F$_y$,~G$_z$) & disordered & 0.0 & 1.0 & 2.4 \\
\hline
\hline
\end{tabular}
\end{center}
\caption{The total energy differences in meV per atom with respect to the ground-state of NdMnO$_3$ for the 
four plausible magnetically ordered states ($\Gamma_{i\mathrm{Mn}}\Gamma_{j\mathrm{Nd}}$, see Tables~\ref{magnSpaceGroup} and SM1 
for definitions of the states) and for the p$_{\rm{AF}}$ (antiferromagnetic Mn$-$Mn), p$_{\rm{F}}$ (ferromagnetic Mn$-$Mn), 
and p$_{\rm{exp}}$ (experimental) lattice structures, see text for details.}
\label{magorder4}
\end{table}

At this point, magnetically ordered options remain, $\Gamma_{\mathrm{5Mn}}\Gamma_{\mathrm{5Nd}}$ and 
$\Gamma_{\mathrm{5Mn}}\Gamma_{\mathrm{Nd}}$, which have very similar total energies for 
a fixed geometry as in experimental studies (p$_{\rm{exp}}$ structure) and for the ferromagnetically 
antiferromagnetically ordered Mn$-$Mn options, p$_{\rm{AF}}$ and p$_{\rm{F}}$ structures, respectively. 
However, the last row in Table~\ref{magorder4}, with magnetic 
ordered state $\Gamma_{\mathrm{5Mn}}\Gamma_{\mathrm{Nd}}$, requires special attention 
since a ``randomly'' oriented $\mu_{\mathrm{Nd}} = 1.4~\mu_{\mathrm{B}}/$atom was used.  
To improve the ``randomness'', 
the simulation window was increased by a factor of two in every dimension to give  
a $2 \times 2 \times 2$ supercell containing 160 atoms, of which 16 are Nd atoms. (Periodic boundary conditions 
means that only magnetic moments inside the supercell are really 
random). This supercell approach did not change any of the details of the calculation 
(\emph{e.g.}~stability and magnetic ordering of the Mn atoms), and now the average Nd magnetic 
moment was closer to zero, as the statistics were significantly improved.  
Therefore, these calculations cannot unambiguously determine if Nd ions order, or not, leaving both of these possibilities acceptable from the theoretical point of view.

\section{Discussion \label{discussion}}
The NdMn$_{0.8}$Fe$_{0.2}$O$_3$ compound mixes an orthomanganite and orthoferrite with similar structures, 
excepting the Jahn-Teller long bond of the Mn.  Magnetically, NdMnO$_3$ is highly anisotropic with A-type 
antiferromagnetism\cite{Munoz2000,Chatterji2009} and NdFeO$_3$ is weakly anisotropic with G-type 
antiferromagnetism.\cite{Belov1972,Koshizuka1988}  Our study is an investigation of single-ion doping in the anisotropic-A-type, 
pseudo-isotropic-G-type phase diagram to better understand the experimental magnetic structure.

The magnetic structure of NdMn$_{0.8}$Fe$_{0.2}$O$_3$ was unambiguously identified to be $\Gamma_5$, (A$_x$, F$_y$, G$_z$) for the Mn ions over the whole temperature range 1.6~K~$\leq T \leq~ T_{\rm N}$.  This structure is within the magnetic space group of NdMnO$_3$ and the high temperature magnetic structure of NdFeO$_3$. On the other hand in low temperature magnetic structure of NdFeO$_3$, G-type antiferromagnetism is accommodated by the $y$-direction, which means the $\Gamma_3$ representation. The most surprising finding of this work is that the magnetic structure of Nd sublattice is also within the $\Gamma_5$, (0, f$_y$, 0) representation and the Nd ions exhibit long range magnetic order at temperatures below $T_{\rm N}$. This finding is different from NdMnO$_3$, where the ordering of Nd ions was reported only below $T_1 \approx 20$~K\cite{Chatterji2009} and different also from NdFeO$_3$, where the ordering of Nd sublattice is suppressed below 4.5~K.\cite{Przenioslo2006} On the other hand, Chatterji {\it et al.} \cite{Chatterji2009bis} conclude from their neutron backscattering data that the ``finite energy of the inelastic peak and its much smaller temperature dependence at $T > 20$~K (are) due to the polarization of the Nd magnetic moment by the field of Mn moments''. In fact, the finite energy of the inellastic peak in NdMnO$_3$ was observed below 40~K. A similar effect was also observed by our neutron backscattering experiment performed on NdMn$_{0.8}$Fe$_{0.2}$O$_3$ below $T_{\rm N}$.\cite{daniel2018} The backscattering results prove the Nd ions become polarized at $T_{\rm N}$. Since the backscattering experiment probes events with characteristic timescale $\sim 10^{-9}$~s, this experiment can not distinguish polarized Nd ions in short range magnetic correlations from those in static long range magnetic structure. Consequently, the Nd ions order at $\approx 20$~K in case of NdMnO$_3$,\cite{Chatterji2009,Chatterji2009bis} whereas the Nd ions order at $T_{\rm N}$ in NdMn$_{0.8}$Fe$_{0.2}$O$_3$, even though both compounds exhibit essentially the same neutron backscattering spectra.

Our analysis shows that the effect observed at $T_1\approx 13$~K by AC susceptibility is the spin reorientation effect. Such an effect was not observed in NdMnO$_3$, but spin reorientation is well reported for NdFeO$_3$ compound.\cite{slawinski2005} Presumably the Fe ions start to destabilize the magnetic structure of the Mn sublattice at the concentration studied in this work ($x = 0.2$), and the spin reorientation is the consequence of Fe doping. Another consequence is the stabilization of the long range magnetic structure of Nd ions.

Finally, in all of the diffraction data sets, there is one sharp reflection for a given family of planes.  
However, as the crystal and magnetic structures of the parent compounds are so similar, it is possible that minor chemical inhomogeneities exist over nanometer-sized length scales.\cite{shulyatev2009}  Along with subtleties of stoichiometry, such effects may be 
important when comparing samples from different laboratories.

\section{Conclusions \label{conclusion}}
The magnetic structure of NdMn$_{0.8}$Fe$_{0.2}$O$_3$ has been investigated using NPD. The resulting model for $T < T_{\rm N}$ has wavevector ${\bf k} = (000)$ and the $\Gamma_5$ magnetic structure with the (A$_x$, F$_y$, G$_z$) configuration for the Mn ions and the (0, f$_y$, 0) arrangement for the Nd ions. The magnetic structure follows the dominant Mn ion, but finds a way to accommodate the interactions of the less populous Fe ion which affects fine details of the magnetic structure. Quantitative analysis to substantiate this model are underway with additional probes. 

\begin{acknowledgments}
This research project has been supported, in part, by the European Commission under the 7th Framework Programme through the 
`Research Infrastructure' action of the `Capacities' Programme, NMI3-II Grant number 283883, VEGA project number 2/0132/16, 
and ERDF EU under the contract No.~ITMS-26220120047, by the US National Science Foundation through Grants DMR-1202033 (MWM) 
and DMR-1157490 (NHMFL), and by the Czech Science Foundation project 14-08124S (DK). 
A portion of this research used resources at the High Flux Isotope Reactor (HFIR), a Department of 
Energy Office of Science User Facility operated by the Oak Ridge National Laboratory.  
The intensive numerical calculations (by DL and KLM) were supported by The 
Ministry of Education, Youth and Sports from the Large Infrastructures for 
Research, Experimental Development and Innovations project „IT4Innovations 
National Supercomputing Center – LM2015070 and National Programme of 
Sustainability (NPU II) project ``IT4Innovations excellence in science'' - LQ1602 
and D.L. also by the Grant Agency of the Czech Republic, project No.  
17-27790S.
\end{acknowledgments}

%

%
%%%%%%Appendix starts here. \newpage command only if needed %%%%%%%%%%%%%%%%%%%%%%%%%%%%%%%%%%%%%%%%%
%
%\newpage
\section*{Appendix A}
\renewcommand\thetable{A1}    
\setcounter{table}{0}  
\begin{table}[h!]
\begin{center}
\begin{tabular}{ccc}
\hline
&Mn(4$b$)&	Nd(4$c$)\\
\hline
$\Gamma_1$&		(G$_x$, C$_y$, A$_z$)&	( -- , c$_y$, -- )\\
$\Gamma_2$& 	--&		(g$_x$, -- , a$_z$)\\
$\Gamma_3$&		(C$_x$, G$_y$, F$_z$)&	(c$_x$, -- , f$_z$)\\
$\Gamma_4$&		--&		( -- , g$_y$, -- )\\
$\Gamma_5$&		(A$_x$, F$_y$, G$_z$)&	( -- , f$_y$, -- )\\
$\Gamma_6$&		--&		(a$_x$, -- , g$_z$)\\
$\Gamma_7$&		(F$_x$, A$_y$, C$_z$)&	(f$_x$, -- , c$_z$)\\
$\Gamma_8$&		--& 		( -- , a$_y$, -- )\\
\hline
\multicolumn{3}{c}{
\begin{tabular}{cc}
F = m$_1$ + m$_2$ + m$_3$ + m$_4$&	C = m$_1$ $-$ m$_2$ + m$_3$ $-$ m$_4$\\
G = m$_1$ $-$ m$_2$ $-$ m$_3$ + m$_4$& 	A = m$_1$ + m$_2$ $-$ m$_3$ $-$ m$_4$\\
\end{tabular} }\\
\hline
\end{tabular}
\end{center}
\caption{Basis vectors for space group $Pnma$ (No. 62 in International Tables for Crystallography\cite{ITC}) and ${\bf k} = (000)$. Atomic positions for Mn: Mn1 $(0, 0, \frac{1}{2})$, 
Mn2 $(\frac{1}{2}, 0, 0)$, Mn3 $(0, \frac{1}{2}, \frac{1}{2})$, Mn4 $(\frac{1}{2}, \frac{1}{2}, 0)$; for Nd: Nd1 $(x, \frac{1}{4}, z)$, 
Nd2 $(-x+\frac{1}{2}, \frac{3}{4}, z+\frac{1}{2})$, Nd3 $(-x, \frac{3}{4}, -z)$, Nd4 $(x+\frac{1}{2}, -y+\frac{1}{2}, -z+\frac{1}{2})$.\label{magnSpaceGroup}}
\end{table}

\end{document}